# Detecting mechanical loosening of total hip replacement implant from plain hip radiograph using convolutional neural network

Alireza Borjali, PhD [a,b*], Antonia F. Chen, MD/MBA[c] , Alisha Sodhi [a], Martin Magneli, MD, PhD [b,d], Kartik M. Varadarajan, PhD [b], Orhun K. Muratoglu, PhD [a,b]

[a] Department of Orthopaedic Surgery, Harris Orthopaedics Laboratory, Massachusetts General Hospital, Boston, MA
[b] Department of Orthopaedic Surgery, Harvard Medical School, Boston, MA
[c] Department of Orthopaedic Surgery, Brigham and Women's Hospital, Harvard Medical School, Boston, MA
[d] Karolinska Institutet, Department of Clinical Sciences, Danderyd Hospital, Stockholm, Sweden

* Corresponding author
Alireza Borjali
Address: 55 Fruit St. GRJ 1121B
Boston, MA, 02114
Email: aborjali@mgh.harvard.edu
a.borjali@gmail.com

**Running Title:** Hip mechanical loosening detection CNN

**Author Contributions Statement:** AB, AC, KV, and OM designed the study. AS and MM collected, cleaned, and labeled the dataset. AF labeled the dataset as the human expert. AB performed the study. AB, KV, AC, and OM performed the analysis and interpretation of the results. Each author has contributed to the writing and revising the manuscript. All authors have read and approved the final submitted manuscript.

## Abstract

Plain radiography is widely used to detect mechanical loosening of total hip replacement (THR) implants. Currently, radiographs are assessed manually by medical professionals, which may be prone to poor inter- and intra-observer reliability and low accuracy. Furthermore, manual detection of mechanical loosening of THR implants may be impacted by the extent of clinical experience, potentially resulting in delayed diagnosis. In this study, we present a fully automatic approach to detect mechanical loosening of THR implants from plain radiographs using deep convolutional neural networks (CNNs). We trained multiple CNNs on 236 cementless THR patients' anteroposterior hip x-rays using a transfer learning (TL) approach and compared their performances with a board-certified orthopaedic surgeon. Mechanical loosening was defined as THR that were surgically revised for loosening of stem and/or acetabular cup and confirmed intra-operatively. To increase the confidence in the machine's outcome, we also implemented saliency maps to visualize where the CNNs focused at to make a diagnosis. The deepest CNN (DenseNet-201) outperformed the other CNNs, achieving 75.0% specificity and 91.6% sensitivity. The saliency maps showed that the CNN focused at clinically relevant features to make a diagnosis. The orthopaedic surgeon achieved 100% specificity and 83.3% sensitivity.  A deep CNN can be trained using the TL approach for automatic radiologic assessment of mechanical loosening of THR implants achieving performance approaching that of an expert orthopaedic surgeon. Such CNN may in the future be used by clinicians to supplement their decision-making process, increase their diagnostic accuracy, and free them to engage in more patient-centric care.

## 1. INTRODUCTION

Plain radiography remains the primary imaging modality used for diagnosing mechanical (aseptic) loosening of total hip replacement (THR) implants due to its low cost, widespread availability, and low-radiation exposure. The basis of radiological assessment of cementless THR implant mechanical loosening is the visual identification of radiolucent regions around the bone-prosthesis interface (periprosthetic lucency). The areas of interest are typically the DeLee and Charnley zones adjacent to the acetabular component, and Gruen zones adjacent to the femoral component [1, 2] (Fig. 1). For the acetabular component of a cementless THR implant, migration or periprosthetic lucency present in all three DeLee and Charnley zones that appears or progresses after 2 years or is > 2 mm in any zone is considered indicative of loosening [1, 2]. For the femoral stem of a cementless THR implant, endosteal scalloping (focal lesions) and a change in position of the implant, including migration and progressive subsidence, as well as lucency regions > 2 mm present in Gruen zones are considered indicative of loosening [3, 4].

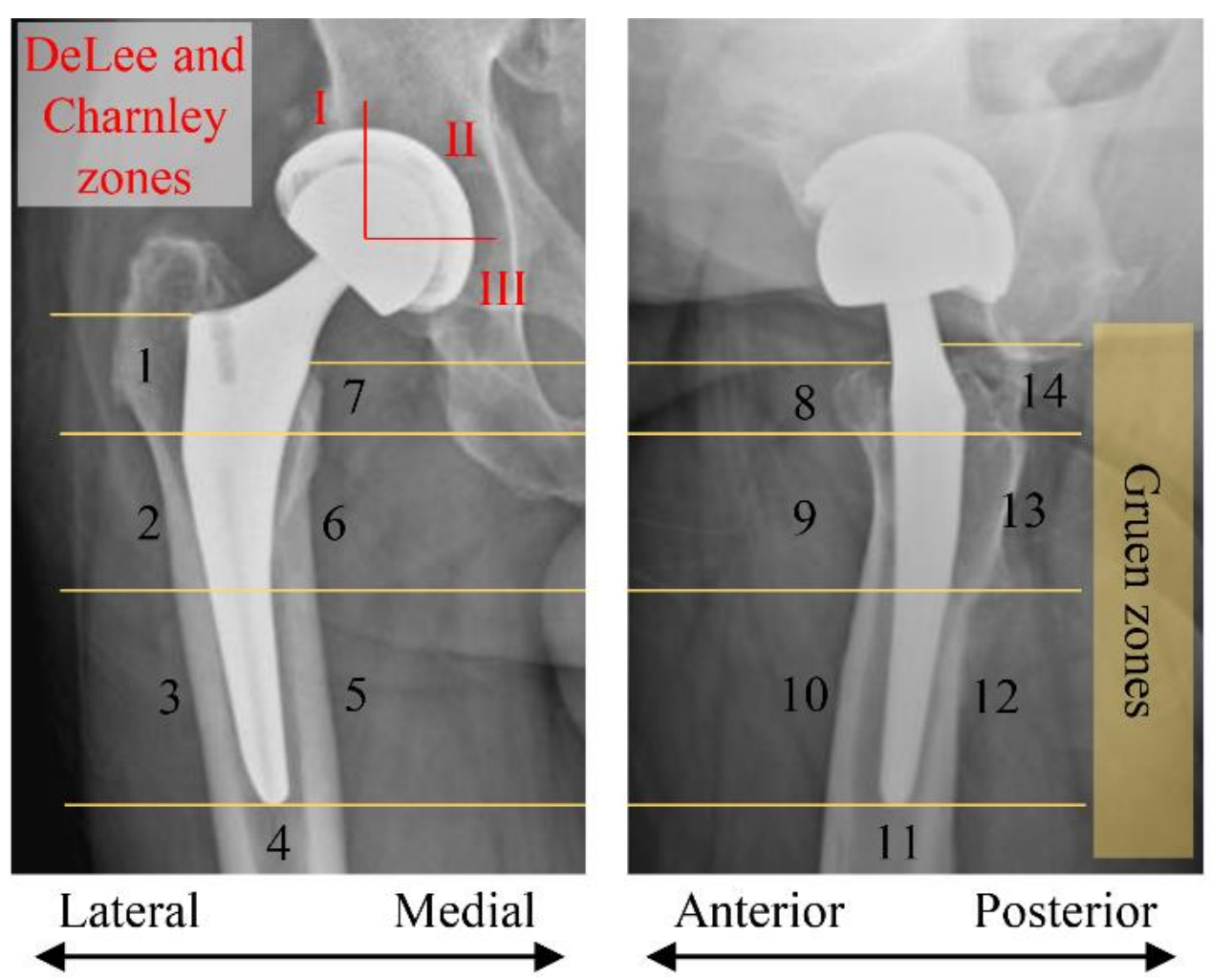

**Fig.1** DeLee and Charnley zones (I,II, and III) adjacent to the acetabular component, and Gruen zones (1-14 ) adjacent to the femoral component.

Currently, radiographs are assessed manually, which may be prone to poor inter- and intra-observer reliability and low accuracy. Individuals identify patterns differently and assign different levels of importance to specific features based on their own experiences. This is evident in the wide range of sensitivities, ranging from 83 to 100%, and specificities, ranging from 55 to 82% in diagnosing mechanical loosening of cementless THR implants that have been reported in the literature [3, 5–8]. Recent breakthroughs in deep learning (DL) algorithms have significantly improved a machine's ability to assess imaging datasets automatically. DL algorithms have already been applied to plain film radiographs with high degrees of success in different orthopaedic applications, such as identification of wrist, elbow, humerus, ankle and hip fractures, classification of the proximal humerus and hip fracture types, detecting presence and type of arthroplasty, and staging knee osteoarthritis (OA) severity [9, 10, 19–21, 11–18], as well as other image dependent diagnosis such as skin maladies [22–24]. Performance of machines in these applications has been typically on-par with trained surgeons and radiologists and superior to general practitioners. DL algorithms can discover complex patterns, which may be nearly impossible for a human observer to compute mentally. Furthermore, machines can learn from medical professionals' collective experiences and consistently apply them to every new patient. With this background in mind, we hypothesized that a deep convolutional neural network (CNN) model could be trained to provide an automated radiographic assessment of mechanical loosening of THR implants.

## 2. METHODS

### 2.1 Study Design

After acquiring institutional review board (IRB) approval, we conducted a retrospective study using previously collected radiographs of THR patients. This study aimed to develop a DL model for diagnosing THR implant mechanical loosening from a given radiograph. We also compared the DL model's performance with an experienced orthopaedic surgeon performing the same task.

### 2.2 Data

The study cohort consisted of previously collected imaging data from 16 centers in 8 countries. We evaluated a total of 15,277 THR patients. First, we identified the patients undergoing a revision THR surgery for any reason using Current Procedural Terminology (CPT) codes (n=9198). We then used the International Classification of Diseases, 10th Revision (ICD-10) codes to include patients who underwent revision THR surgery only due to mechanical loosening of a cementless THR implant (n=316). Any patient with multiple codes (e.g. mechanical loosening + fracture) was excluded. Subsequently, we evaluated the operative notes of these patients. We only included patients with intra-operative confirmation of THR mechanical loosening with no sign of infection diagnosed during surgery by the orthopaedic surgeon and reported in the operative note (n=118). Any patient with differing pre-operative and post-operative diagnoses was excluded. The immediate (up to 1 month) pre-revision THR surgery anterior posterior (AP), non-weight bearing, hip radiographs of these patients were collected. These radiographs were labeled as "mechanically loose". It is important to reiterate that these labels were surgically confirmed, creating a solid ground truth for DL models.

We did not differentiate between acetabular and femoral component loosening. Table 1 summarizes all the CPT and ICD-10 codes that were used to select the patient cohort.

**Table 1** Current Procedural Terminology Code (CPT) and International Classification of Diseases, 10th Revision (ICD-10) codes indicating primary and revision hip arthroplasty and hip prosthetic joint mechanical loosening diagnosis

| Type | Code | Description |
|---|---|---|
| *CPT | 27130 | Arthroplasty, acetabular and proximal femoral prosthetic replacement with or without autograft or allograft |
| | 27134 | Revision of total hip arthroplasty; both components, with or without autograft or allograft |
| | 27137 | Revision of total hip arthroplasty; acetabular component only, with or without autograft or allograft |

| | 27138 | Revision of total hip arthroplasty; femoral component only, with or without allograft |
|---|---|---|
| **ICD-10 | T84.031A (D/S) | Mechanical loosening of internal left hip prosthetic joint, initial (subsequent/ sequela) encounter |
| | T84.030A (D/S) | Mechanical loosening of internal right hip prosthetic joint, initial (subsequent/ sequela) encounter |

*CPT: Current Procedural Terminology Code
**ICD-10: International Classification of Diseases, 10th Revision

A comparison group from the same study cohort was developed comprised of 118 patients. The matching criteria was primary cementless THR surgery. These patients had no reported post-operative complications with a minimum of one-year follow-up. The AP, non-weight bearing, hip radiographs of these patients taken at the one-year time point were collected and were labeled as “well-fixed”. Figure 2 shows the flow diagram of included patients.

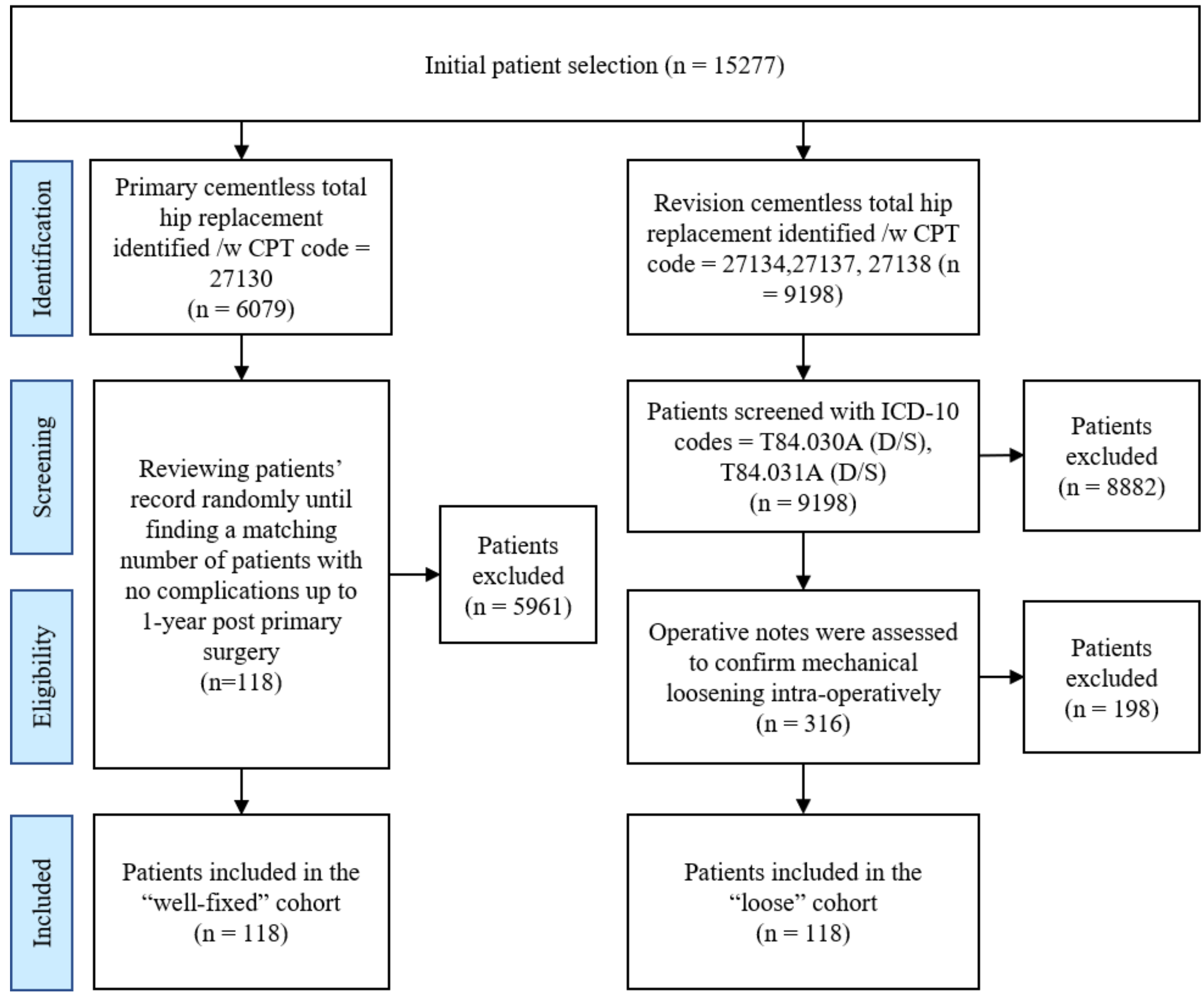

**Fig. 2** Flow diagram of included patients

Figure 3 shows examples of "well-fixed" implant radiograph (Fig. 3 [a]) and "loose" implant radiograph indicating the loosening regions (Fig. 3 [b]).

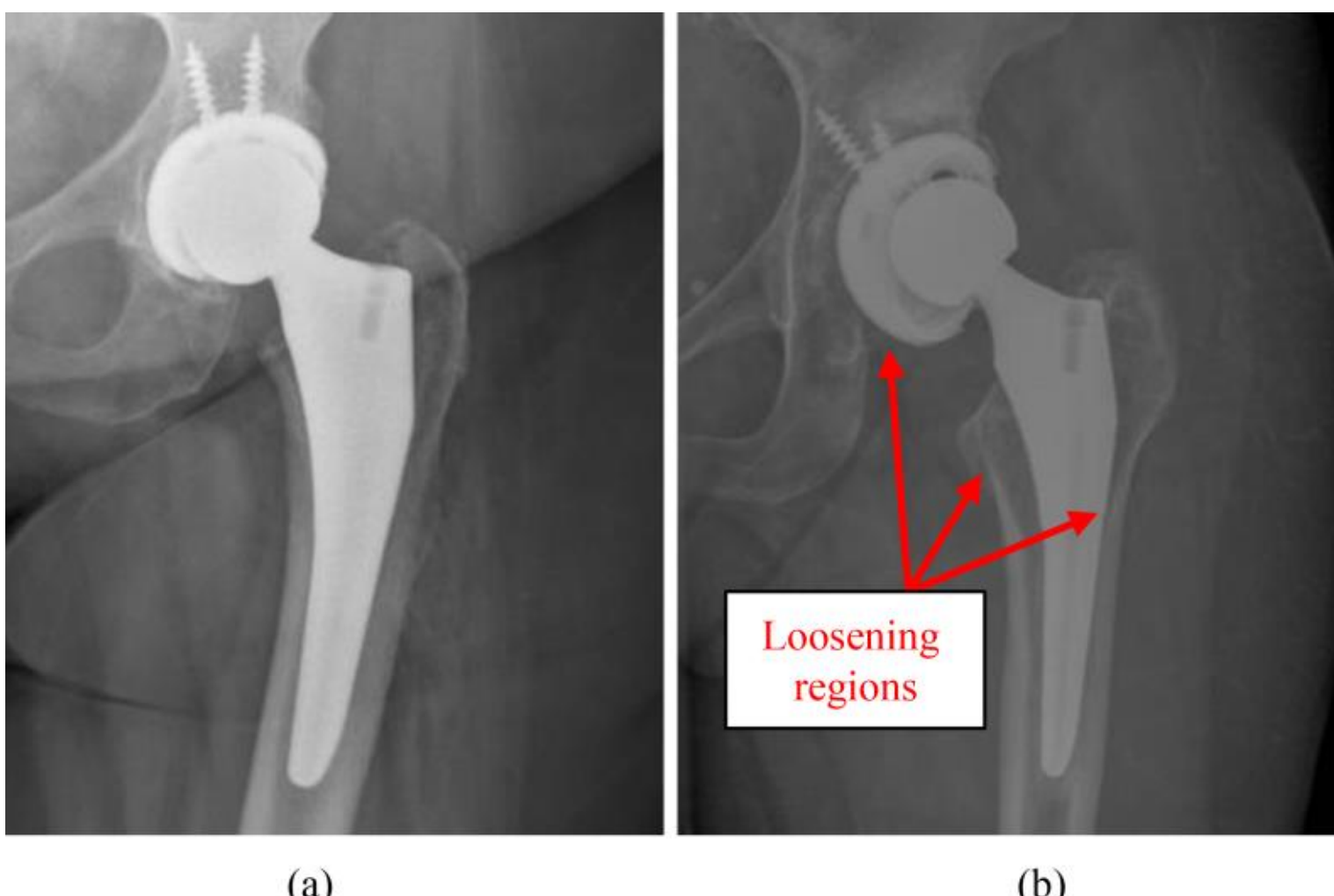

**Fig. 3** Examples of (a) well-fixed implant radiograph and (b) loose implant radiograph indicating the loosening regions.

## 2.3 Data Preparation

Some radiographs were in JPEG format and some were in digital imaging and communications in medicine (DICOM) format. In the first step, all the embedded patient information in the DICOM files was removed. Subsequently, these radiographs were also converted to JPEG format using MicroDicom Viewer software (MicroDicom Viewer, version 3.0.1). In the next step, all radiographs were rescaled to fit the DL models' input (224 by 224 pixels) and they were normalized by dividing each pixel's value by 255. The split validation method was implemented using an 80:10:10 split ratio to randomly divide the dataset into training, validation, and final test subsets [25]. Data augmentation was employed on the training subset by applying horizontal flip, variation in magnification, brightness, hip positioning, and hip orientation between 0 to 50%. Data augmentation is a well-established method to develop DL models that reduces the probability of overfitting by increasing the invariance of the model to real-world discrepancies between the radiographs. The validation subset was used for hyperparameter optimization. In the end, the final models were evaluated on the test subset that was isolated from the training and validation process.

## 2.4 Models

A convolutional neural network (CNN) is a commonly used machine learning method for image analysis and classification. Training a CNN from scratch requires a large dataset that is usually a challenge in medical applications. The transfer learning (TL) approach can mitigate the lack of data by transferring and re-using the learned parameters (i.e., network weights) of a well-trained CNN model on a large dataset in one domain to solve a problem in another domain with a smaller dataset. We employed a TL approach with CNNs well-trained on ImageNet dataset [21]. ImageNet dataset contains over 15 million non-medical, labeled images. We chose the most successful CNNs in classifying images in the ImageNet dataset. Using the TL approach with these CNNs have shown promising results for medical image analysis in recent years [19, 20, 26]. These CNNs are briefly explained in the following subsections and more details can be found elsewhere [21, 27].

### *2.4.1 VGGNet*

Visual Geometry Group (VGG) first introduced VGG-16 in 2014 followed by VGG-19 as two successful architectures classifying ImageNet dataset. These models use a large kernel-sized filters with multiple small kernel-sized filters resulting in 13 and 16 convolution layers with 3 fully connected layers: hence the name VGG-16 and VGG-19 respectively. We implement VGG-19 which is the improved version of VGG-16 [28].

### *2.4.2 Inception*

Inception V1 (also called GoogLeNet) introduced by Google Inc. tried to improve the memory efficiency and runtime of VGGNet while maintaining the performance. Inception replaces the activation functions of VGGNet that are redundant or zero with a module called Inception that approximates sparse connections between the activation functions.  Inception-V1 was further refined in three other versions. Inception-V2 used batch normalization, V3 proposed a factorization method, and Inception V-4 introduced a uniform simplified version of Inception-V3. We implemented Inception-V3 in this study [29].

*2.4.3 ResNet*

ResNet CNN introduced by Microsoft Research implemented Residual Learning to tackle accuracy saturation and vanishing gradients problem of deep networks. Unlike other CNNs that learn features, ResNet learns the subtraction of learned features from the input for each convolution layer (residuals). There are different versions of ResNet with different number of layers (e.g., ResNet-34, ResNet-50, and ResNet-101). We implemented ResNet-50 [30].

*2.4.4 Inception-Residual Network*

This CNN model combines the Inception and ResNet architectures. On the one hand, Inception can learn features at different resolutions, and on the other hand, ResNet enables the network to have a deeper architecture to learn more complex features. We implemented Inception-ResNet-V2 [23] that is based on Inception-V4 [31].

*2.4.5 Xception*

Xception (extreme inception) is a modified version of the Inception-V3 using depth-wise separate convolutions for spatial dimension and channel dimension of a given image during the training process [32].

*2.4.6 DenseNet*

Densely Connected Convolutional Network (DenseNet) modifies the connectivity pattern between different convolution layers. In the DenseNet architecture, each convolution layer receives the output of all preceding layers mitigating the vanishing gradient problem as well as decreasing the number of feature maps. DensNet has different versions (e.g., DenseNet-121, DeneNet-169, and DenseNet-201). We implanted DenseNet-201 in this study [33].

### 2.5 Training

We replaced the classifier layers of all CNNs with five layers as follows: two layers of fully connected neural network (dense layers) with 128 and 32 neurons respectively, followed by a dropout (0.3) layer to further reduce the chance of overfitting, followed by one output neuron for binary classification of x-rays into “well-fixed” or “loose”. We used a fine-tuning training approach by leaving the convolution layers of the well-trained CNNs unfrozen and their weights could get updated during the training process. We initialized the classifier layers with random weights. We trained the CNNs using Adam optimizer, for 10 epochs, with a batch size of 16, and an initial learning rate of 0.00001. We implemented CNN using Tensorflow (Keras) on a workstation comprised of an Intel(R) Xeon(R) Gold 6128 processor, 64GB of DDR4 RAM, and an NVIDIA Quadro P5000 graphic card.

### 2.6 Evaluation

#### *2.6.1 Metrics*

We calculated the sensitivity and specificity of the CNNs and plotted the receiver operating characteristic (ROC) curve. Sensitivity (true positive rate) was the ratio of correctly identified loose x-rays by CNNs to all of the loose x-rays in the test subset, and specificity (true negative rate) was the ratio of correctly identified well-fixed implant x-rays by CNNs to all the well-fixed implant x-rays in the test subset.

#### *2.6.2 Visualization*

We implemented image-specific saliency maps to indicate the importance of each pixel of a given x-ray on the CNN's classification. Saliency maps helped to visualize the CNN process and indicated where the network was focusing at to make a classification [34]. Saliency maps shed light on the CNN decision-making process and increased confidence in its outcome.

### *2.6.3 Human expert comparison*

We compared the CNNs' performance with human expert performance in diagnosing THR implant mechanical loosening. For this, we provided a board-certified orthopaedic surgeon with the same information as the CNN. This human expert was only shown the x-ray images in a blinded fashion without access to any other information about the patients, such as medical history or additional radiographs. We measured human expert's sensitivity and specificity on the test subset.

## 3. RESULTS

Figure 4 summarizes the ROC curves over the test subset for all CNNs classifying the x-rays into "loose" and "well-fixed" implant categories. Figure 5 shows that the shallowest CNNs (VGG-19 and ResNet50) achieved the lowest overall area under the ROC curve (AUC = 0.62). Inception-V3 and Xception achieved similar and slightly higher AUC (0.77). This is in line with the results in ImageNet, since Inception-V3 and Xception achieved similar results in that dataset as well [32]. Inception-ResNet-V2 achieved higher AUC (0.80) in comparison with both ResNet-50 and Inception-V3 combining the strengths of both CNNs. DenseNet-201, which was the deepest CNN in this study, achieved the highest overall AUC (0.84).

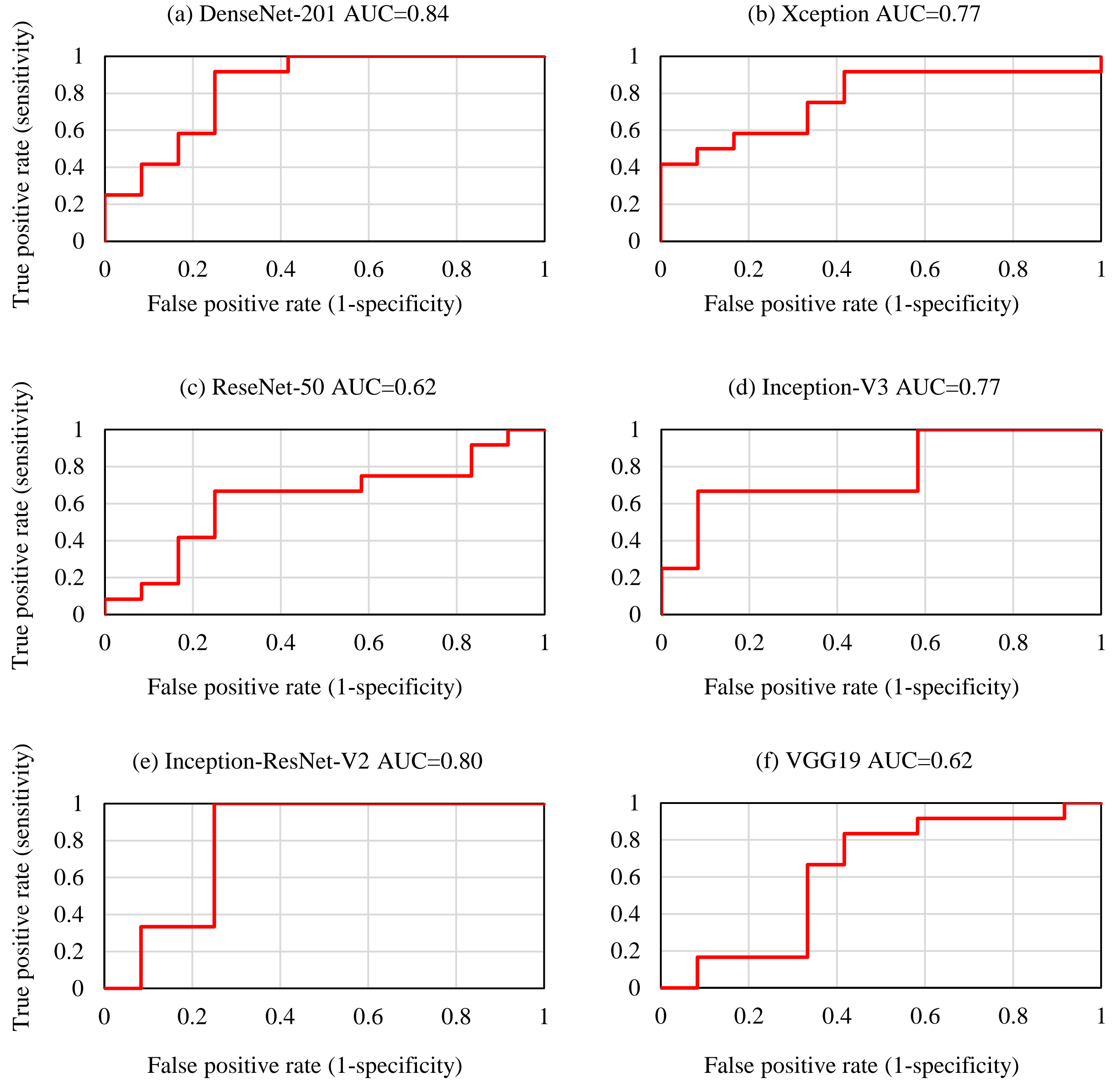

**Fig. 4** Receiver operating characteristic (ROC) curve with area under curve (AUC) over the test subset for (a) DenceNet-201, (b) Xception, (c) ResNet-50, (d) Inception-V3, (e) Inception-ResNet-V2, and (f) VGG19 CNNs.

The orthopaedic surgeon achieved 100% specificity and 83.3% sensitivity. The orthopaedic surgeon made two overall mistakes misdiagnosing two "loose" patients as "well-fixed". The high specificity of the orthopaedic surgeon indicates that the surgeon could identify the well-fixed THR implants perfectly but made two errors in diagnosing the loose x-rays. The best performing CNN (DenseNet-201) achieved 75.0% specificity and 91.6% sensitivity at a threshold selected from Fig. 4.

Figure 5 shows example x-rays and the corresponding saliency map using best performing CNN (DenseNet-201) for both loose and well-fixed cases. Colored regions in the saliency maps indicated the most influential regions on the CNN's performance, where red denoted higher relative influence than blue. The saliency maps identified the significant influence of bone-implant interaction regions on the CNN's performance. This was in line with the clinical diagnosis of THR implant mechanical loosening, since the CNN was 'looking' at the same clinically relevant regions in the x-rays as a radiologist or an orthopaedic surgeon would look at to make a diagnosis.

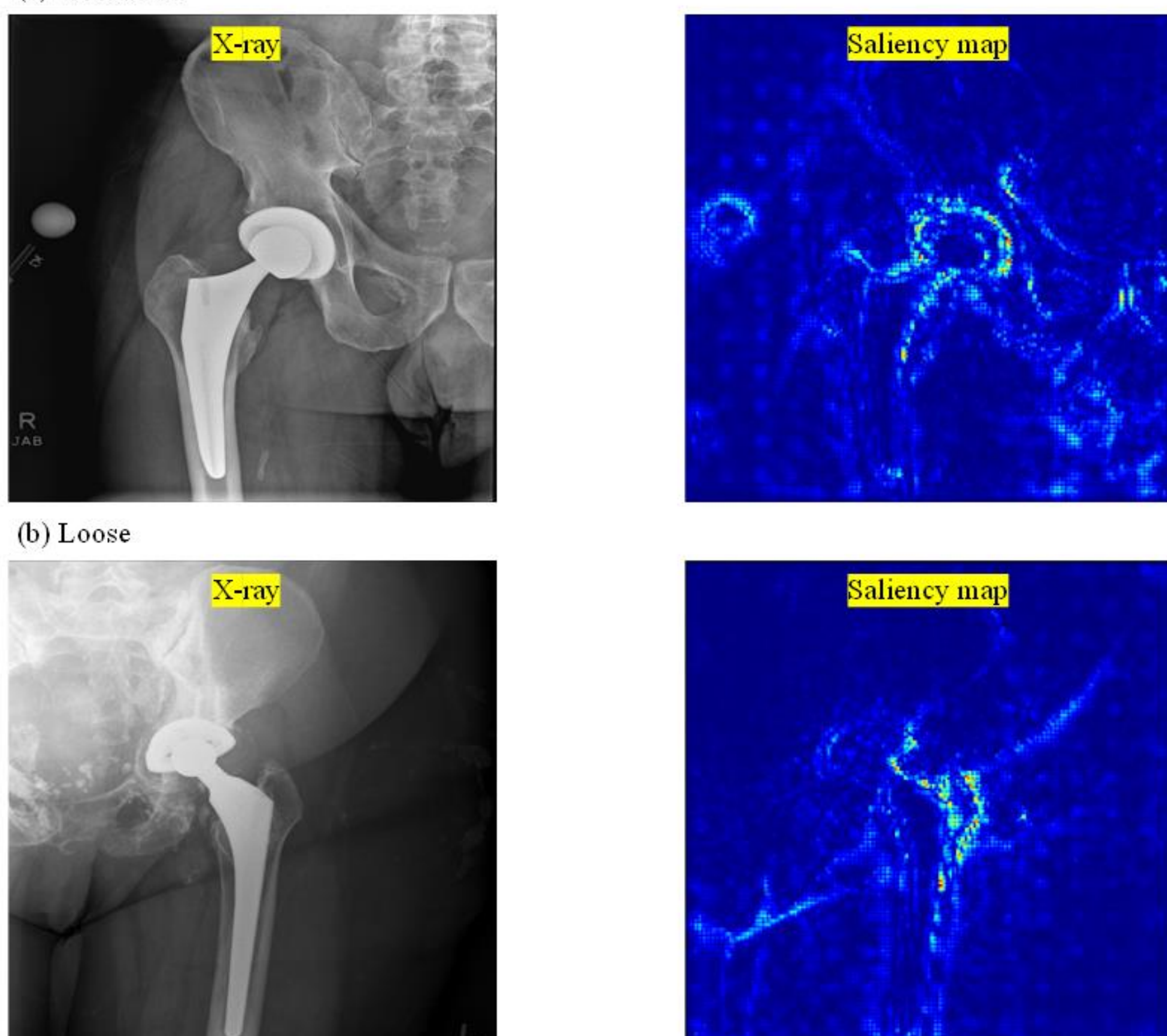

**Fig. 5** X-rays and saliency maps of best performing CNN model (DenseNet-201) for example (a) well-fixed, and (b) loose patients. Colored regions, where red denotes higher relative influence than blue, indicate most influential regions on the CNN's performance.

## 4. DISCUSSION

We implemented CNN models using a TL approach to automatically detect mechanical loosening of THR implants from plain film radiographs. We benchmarked different CNN models against each other and the manual radiographic assessment of an expert orthopaedic surgeon. Other studies have applied DL algorithm on plain film radiographs for various orthopaedic applications [9–17]. However, to the best of our knowledge, this is the first study to apply DL algorithm for the automatic detection of mechanical loosening of THR implants.

Furthermore, we compared the performance of different CNN architectures in performing the same medical task. In a recent review of the literature, we identified this issue as a knowledge gap that in most cases medical studies do not benchmark the CNN model that they use and only implement one model with little to no reasoning [21]. Finding the optimal model for a specific medical task can only be achieved by benchmarking different models against each other on the same dataset. A stronger focus on systematic benchmarking among all orthopaedic artificial intelligence (AI) studies through standardized methods would be critical to determine optimal models for orthopaedic applications.

We showed that the deepest CNN (DenseNet-201) achieved the best overall performance among other CNNs. Furthermore, DenseNet-201 achieved higher sensitivity and lower specificity than the orthopaedic surgeon. Other studies have reported a wide range of sensitivities from 83 to 100%, and specificities from 55 to 82% for manual radiographic assessment of mechanical loosening of THR implants [5-8]. This shows the potential of integrating the CNN model that we developed in orthopaedic care to potentially increase diagnostic accuracy and consistency. We have also used saliency maps to shed light on the decision-making process of the CNN. We showed that the CNN looked at clinically relevant features to make a diagnosis of mechanical loosening in THR. This visualization is critical to build confidence in the machine's outcome and move towards the integration of AI in daily orthopaedic care.

The primary limitation of this study is the size of the dataset. Although analysis akin to traditional statistical tests are not available for CNN models, it is expected that with larger datasets in future studies, we can increase CNNs' performance and improve their reliability. Nonetheless, the TL approach employed herein is a validated method for developing CNN models on relatively small datasets. We further compensated for the small size of our dataset by utilizing the data augmentation method. Data augmentation increased the number of training images by applying small changes to the base dataset and created new x-rays. These small changes accounted for small naturally occurring variations in x-ray images, such as different hip positioning and image quality. Furthermore, we were only able to assess cementless THRs. With a larger dataset we can potentially include both cementless and cemented THRs in our study.

Another limitation of this study is that the CNN was trained on late-stage mechanical loosening cases, using radiographs obtained just before revision surgery. Thus, the performance of CNN on detecting early stages of THR implant mechanical loosening is unknown. Expansion of the dataset will also help to overcome this limitation. Furthermore, we only used one AP, non-weight bearing, hip radiographs per patient, as opposed to clinical practice where the clinician would have the benefit of additional views, and comparison to serial radiographs to evaluate the progression of the mechanical loosening. Furthermore, we did not compare radiographs to other modalities of detecting mechanical loosening.

In this study, we presented a novel, fully automatic and interpretable approach to detect mechanical loosening of THR implants from AP hip x-ray using deep CNNs and TL approach. These CNNs can be used in clinics to assist physicians in diagnosing THR mechanical loosening and potentially improve their diagnostic accuracy.

**Acknowledgement**

This study was supported by laboratory sundry funds, and no external funding was received. The authors have no financial disclosures relevant to this study.